\documentclass{usenix}

\usepackage[noadjust]{cite}
\usepackage{graphicx}
\usepackage{subfigure}
\usepackage{url}
\usepackage{endnotes}
\usepackage{color}

\usepackage[delayed, graphics, dvips, showlabels,sections,textmath,displaymath]{preview}

\newcommand{\beq}{\begin{equation}}
\newcommand{\eeq}{\end{equation}}
\def\bearn{\begin{eqnarray*}}
\def\eearn{\end{eqnarray*}}
\def\barr{\begin{array}}
\def\earr{\end{array}} 

\def\bt{BitTorrent}

\begin{document}

\title{One Bad Apple Spoils the Bunch:\\\large{\it Exploiting P2P
    Applications to Trace and Profile Tor Users}}

\author{ \authname{Stevens Le Blond \quad Pere Manils \quad Abdelberi
    Chaabane} \and \authname{Mohamed Ali Kaafar \quad Claude
    Castelluccia \quad Arnaud Legout \quad Walid Dabbous}
  \authaddr{I.N.R.I.A, France} }

\maketitle 
\sloppy
\begin{abstract}

Tor is a popular low-latency anonymity network.  However, Tor does not
protect against the exploitation of an insecure application to reveal
the IP address of, or trace, a TCP stream.  In addition, because of
the linkability of Tor streams sent together over a single circuit,
tracing one stream sent over a circuit traces them all.  Surprisingly,
it is unknown whether this linkability allows in practice to trace a
significant number of streams originating from secure (i.e., proxied)
applications.

In this paper, we show that linkability allows us to trace 193\% of
additional streams, including 27\% of HTTP streams possibly
originating from ``secure'' browsers.  In particular, we traced 9\% of
\textit{all} Tor streams carried by our instrumented exit nodes.
Using BitTorrent as the insecure application, we design two attacks
tracing BitTorrent users on Tor.  We run these attacks in the wild for
23 days and reveal 10,000 IP addresses of Tor users.  Using these IP
addresses, we then profile not only the BitTorrent downloads but also
the websites visited \textit{per country of origin of Tor users}.  We
show that BitTorrent users on Tor are over-represented in some
countries as compared to BitTorrent users outside of Tor.  By
analyzing the type of content downloaded, we then explain the observed
behaviors by the higher concentration of pornographic content
downloaded at the scale of a country.  Finally, we present results
suggesting the existence of an underground BitTorrent ecosystem on
Tor.

\end{abstract}

\section{Introduction}
\label{sec:intro}

Assume that a source wants to leak top secret documents anonymously.
It is considered secure to do so through Tor using a privacy-enhancing
browser plugin such as TorButton.  However, assume that, at the same
time, this source uses another insecure application on Tor.  Is it
then possible to associate the top secret documents with the IP
address of the anonymous source?  The answer to this question is yes!

By exploiting Tor's design, one can indeed exploit an insecure
application to associate the usage of a secure application (e.g., the
one leaking top secret documents) with the IP address of a Tor user.
This attack against Tor consists of two parts: (a) exploiting an
insecure application to reveal the source IP address of, or
\textit{trace}, a Tor user and (b) exploiting Tor to associate the
usage of a secure application with the IP address of a user (revealed
by the insecure application).  As it is not a goal of Tor to protect
against application-level attacks, Tor cannot be held responsible for
the first part of this attack.  However, because Tor's design makes it
possible to associate streams originating from secure application with
traced users, the second part of this attack is indeed an attack
against Tor.  We call the second part of this attack the \textit{bad
  apple attack}.  (The name of this attack refers to the saying ``one
bad apple spoils the bunch.''  We use this wording to illustrate that
one insecure application on Tor may allow to trace other
applications.)

This paper differs from the related work in three main aspects.
First, we launched our attacks on the real Tor network for a
substantial period of time and revealed 10,000 IP addresses of
``anonymous'' Tor users.  To the best of our knowledge, it is the
largest attack against the Tor network in number of revealed IP
addresses.

Second, whereas most attacks against Tor were targeted to Web
browsers, we directly target P2P filesharing applications (i.e.,
BitTorrrent) in this study.  BitTorrent traffic generates a
significant fraction of Tor traffic in volume (more than 40\%), making
it a primary target for attackers.  In addition, we show that 70\% of
BitTorrent users on Tor establish P2P connections {\it outside} of
Tor, making most BitTorrent TCP connections (and traffic) invisible to
the Tor network.  Thus, the number of BitTorrent users on Tor is
likely to be largely underestimated and so more Tor users are
susceptible to our attacks.

Third, whereas the principle of the bad apple attack has been
discussed in the past, it is an open question whether it allows to
trace a significant number of streams originating from secure
applications.  Actually, that ''many TCP streams can share one
circuit'' is listed as the fourth {\it improvement} of Tor over the
old onion routing design \cite{Tor} because it is supposed to
``improve efficiency and anonymity.'' We note that Roger Dingledine
cited an initial version of this work \cite{TOR-BT-Poster, TOR-BT-TR}
to confirm the need to ``brainstorm about ways to protect users even
when their applications are handing over their sensitive information''
on the website of the Tor Project \cite{TOR-Blog}.  The main
contributions of this paper are as follow:

\begin{itemize}

\item We design two attacks against BitTorrent to reveal the IP
  address of BitTorrent users on Tor.  

\item We instrument six Tor exit nodes and launch our attacks on the
  real Tor network for a period of 23 days.  We reveal 10,000 IP
  addresses of ``anonymous'' Tor users.

\item We show that the bad apple attack allows us to trace 193\% of
  additional streams as compared to BitTorrent streams, including 27\%
  of HTTP streams.  In total, we traced 9\% of \textit{all} Tor
  streams carried by our instrumented exit nodes.

\item We profile BitTorrent and Web usage on Tor per country of origin
  (which would be impossible without first tracing Tor users).  $(a)$
  We show that \bt{} users on Tor are over-represented in some
  countries as compared to \bt{} users that do not use Tor.  $(b)$ By
  analyzing the type of content downloaded, we explain this behavior
  by the higher concentration of pornographic content downloaded at
  the scale of a country.  $(c)$ We present results that suggest the
  existence of an underground \bt{} ecosystem on Tor.

\end{itemize}

The rest of this paper is organized as follow.  First, we briefly
discuss the ethical and legal considerations of running attacks
against production systems in Section~\ref{sec:ethics}.  We then give
an overview of Tor and explain how this design can lead to the bad
apple attack when an insecure application reveals the source IP
address of a Tor user in Section~\ref{sec:onion}.  In
Section~\ref{sec:attacks}, we discuss our attack model, two attacks
against \bt{}, and how the bad apple attack specifically applies to
\bt{}.  We show that the bad apple attack is severe enough to profile
not only \bt{} downloads but also the websites visited by Tor users in
Section~\ref{sec:profil}.  We discuss the related work in
Section~\ref{sec:related}.  Finally, we summarize our contributions
and give a few perspectives in Section~\ref{sec:discuss}.

\section{Ethical and Legal Considerations}
\label{sec:ethics}

In order to comply with the legal and ethical aspects of privacy, we
performed our analysis on the fly and do not store any nominative
information such as IP addresses.  We logged only the ASN and country
of traced Tor users to be able to perform this study.  In addition, we
only present aggregated statistics as suggested by Loesing et
al. in~\cite{loesing:financialcrypto2010}.  Finally, we have also been
cautious not to inadvertently DoS Tor or \bt{} infrastructures, or
interfere with the normal usage of those systems.

\section{Background}
\label{sec:onion}

\subsection{Tor}

Tor is a low-latency anonymity network. As stated in the original
paper, its main design goals are to prevent attackers from linking
communication partners and from linking multiple communications to or
from a given user. Tor relies on an overlay network and on onion
routing to anonymize TCP-based applications like web browsing and P2P
filesharing.  Tor explicitly made the design choice to support {\it
  only} TCP which ``helped portability and deployability'' \cite{Tor}.

When a client communicates with a server via Tor, she selects $n$
nodes of the Tor system (where $n$ is typically 3) and builds a
circuit using those nodes.  Messages are then encrypted $n$ times,
first with the key shared with the last node (called \textit{exit
  node}) of the circuit, and subsequently with the shared keys of the
intermediate nodes from $node_{n-1}$ to $node_1$.  As a result, each
intermediate node only knows its predecessor and successor, but no
other nodes of the circuit.  In addition, only the exit node is able
to recover the original message.

To improve efficiency, Tor multiplexes several streams from the same
source into a single circuit.  Originally, onion routing used a
separate circuit for each stream but it required multiple public key
operations for every request \cite{Reed98anonymousconnections}.  It
has also been argued that creating many circuits degraded privacy
because it implied to contact more Tor nodes, some of which may be
compromised.  However, we will show that, when several streams are
multiplexed into a single circuit, a single stream whose source IP
address is revealed allows an attacker to associate many additional
streams with the same traced user.

\subsection{\bt{}}

\bt{} is a popular Peer-to-peer (P2P) protocol for file replication.
To download a content, a \bt{} client first discovers peers sharing
that content using centralized trackers, a distributed tracker (DHT),
and peer exchange (PEX).

\textit{Trackers} are servers storing content identifiers and for each
identifier, a list of peers distributing the corresponding content.  A
peer subscribes its IP/port to the tracker for a given content
identifier and requests a list of peers for that content when it
starts downloading and then periodically after that (e.g., every 10
minutes).  Communications with trackers is typically done in clear
over TCP (i.e., through Tor) therefore \textit{a malicious exit node
  can tamper with the lists of peers returned by centralized
  trackers.}  (Furthermore, centralized tracking is sometimes also
done over UDP with consequences similar to those we discuss after.)

In addition to centralized trackers, \bt{} clients can also use a
decentralized tracker based on a Distributed Hash Table (\textit{DHT
  tracking}).  Whereas DHT tracking is often used in combination to
centralized tracking, it can also be used alone with Magnet Links.
\bt{} Mainline DHT tracking works as follow.  First, a \bt{} client
(DHT tracker) picks an identifier that is coded on the same space as
content identifiers.  Then, a peer interested in downloading a content
uses the content identifier to locate the corresponding DHT tracker
after which, it subscribes its IP/port to that tracker and requests a
list of peers (just as with a centralized tracker).  Communication
with DHT trackers is done over UDP therefore \textit{a Tor user may
  subscribe her public IP/port to the DHT tracker.}

As a \bt{} client discovers peers, it tries establishing TCP
connections and if successful, sends an application handshake
containing the content identifier and in the case of an extended
handshake, the listening port number.  That P2P connection is also
used for content distribution.  Whether a P2P connection is
established through Tor has a tremendous impact on performance and can
be configured by the user therefore \textit{a Tor user may establish
  P2P connections using her public IP/port.}

Finally, after using centralized or DHT tracking, more peers can be
discover using Peer Exchange (\textit{PEX}).  With PEX, users
typically exchange lists of peers they are connected to over
established P2P connections therefore \textit{a Tor user may subscribe
  her public IP/port to her PEX partners.}

\section{Attacking \bt{} Users on Tor}
\label{sec:attacks}

\subsection{Attack Model}

All our attacks require to control one exit node in order to trace its
Tor users.  From January $15$ to February $7$th 2010 (23 days), we
instrument and monitor six Tor exit nodes spread throughout the world
(two in Asia, two in Europe, and two in the U.S) and launch the
attacks described after.  The first attack, the hijacking tracker's
responses, also requires to control a \bt{} peer publicly connectable
so it can accept incoming TCP connections and receive \bt{} handshake
messages.  For practical reasons, we performed this hijacking attack
on only one of our exit nodes.

Hijacking tracker's responses exploits the fact that centralized
tracking is done through Tor and the list of peers can be tampered
with by a malicious exit node.  One condition for this attack to trace
\bt{} users is that they do content distribution outside of Tor.  We
will see in Section~\ref{sec:hijack} that this is the case for 70\% of
the \bt{} users on Tor.  The second attack exploits the fact that DHT
tracking uses UDP and so is done outside of Tor.  We note that if DHT
tracking was instead done over TCP through Tor, it would still be
possible to perform an hijacking attack as with centralized tracking.
We suspect that it is possible to perform similar statistical attacks
with PEX and centralized UDP tracking, however, we did not exploit
them in this study.

\subsection{Tracing \bt{} Streams on Tor}

\subsubsection{Hijacking Tracker's Responses}
\label{sec:hijack}

\begin{figure}[!t]
  \centering
  \includegraphics[width=0.6\columnwidth]{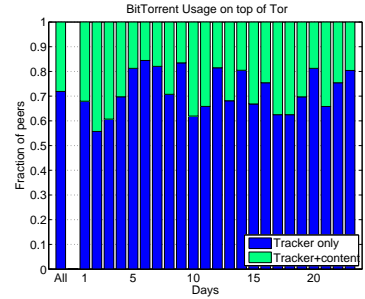}
  \caption{For each day, this histogram shows the proportion of
    BitTorrent peers who use Tor only to connect to the centralized
    tracker (\textit{Tracker only}) or also to distribute content
    (\textit{Tracker+content}).  \textit{All} is the average over all
    days.  \textit{$72\%$ of peers use Tor only to connect to the
      tracker.}}
  \label{fig:bt_usage}
\end{figure}

Hijacking the tracker responses consists in inserting the IP/port of a
peer controlled by the attacker (malicious peer) into the list of
peers returned by the tracker so the targeted user connects to the
malicious peer.  When \bt{} peers use Tor only to connect to the
centralized tracker and not to distribute content, they will connect
to the malicious peer directly, i.e., outside of Tor, allowing the
attacker to trace them.  But how can the attacker distinguish between
a direct connection from one that is going through Tor?

This can be done by collecting the IP addresses of Tor exit nodes
(which are public) to check whether an incoming connection at the
malicious peer originates from one of these addresses.  If it does,
the targeted user is distributing content through Tor.  Otherwise, he
uses Tor only to connect to the tracker and the connection to the
malicious peer is direct (the source IP address field of the IP
datagrams contains the real IP address of the targeted user).  

In Fig.~\ref{fig:bt_usage}, we observe that most \bt{} users use Tor
only to connect to the centralized tracker, making the hijacking of
tracker's responses a simple yet efficient attack.  One explanation
for this behavior may be that users distribute content outside of Tor
to not degrade performance.  In addition, Piatek et al. have showed
that naive spies use tracker subscriptions as evidences of copyright
infringement \cite{BT-DMCA}.  This might be another reason why users
are mainly concerned by anonymizing their tracker subscription.

Because most P2P connections are established {\it outside} of Tor,
most BitTorrent streams (and traffic) are invisible to the Tor
network.  Thus, the number of BitTorrent users on Tor is likely to be
largely underestimated and so more Tor users are susceptible to our
attacks.

\subsubsection{Statistical Exploitation of DHT Tracking}
\label{sec:DHT}

\begin{figure}[!t]
  \centering
  \includegraphics[width=0.6\columnwidth]{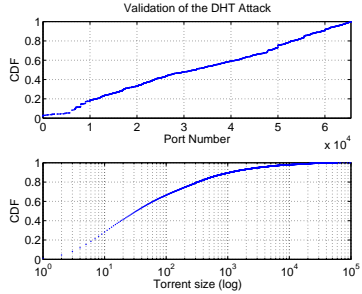}
  \caption{The distribution of the listening port number subscribed to
    the Mainline DHT is uniform (top plot) and most torrents have few
    peers (bottom plot), resulting in a small number of collisions
    among listening ports for peers in the same torrent. \textit{The
      listening port number constitutes a good identifier of a peer
      within a torrent.}}
  \label{fig:dht_valid}
\end{figure}

The second attack exploits DHT tracking.  Because DHT tracking is
carried over UDP (which Tor does not support) a \bt{} client fails to
connect to the DHT through Tor and reverts to using its {\it public}
interface to publish his (public) IP address and listening port into
the DHT.  Say the targeted user is downloading content1.  The IP
address and listening port of that user are then stored on the peer
responsible of tracking content1 in the DHT.  But how can an attacker
in Tor locate the peer storing the public IP address of the targeted
user?  And then, how to distinguish the IP address of the targeted
user from the other IP addresses downloading content1?

To find the information of a targeted user in the DHT, we use the
content identifier and listening port number contained in the \bt{}
subscription to the centralized tracker and extended handshake
messages.  When one of our exit nodes receives one of these messages,
it immediately locates the peer tracking that content identifier and
collects all the IP/port couples that have subscribed for that
identifier.  All peers who have subscribed for that content identifier
are candidates to be associated with the Tor user.  We validate that
the listening port is a good identifier of a peer within a torrent in
Fig.~\ref{fig:dht_valid}.  We then associate the IP address of the
only peer with a matching listening port to the targeted user.  If
there is no such peer or that there is more than one, we consider that
we have failed to trace the targeted user.

\subsection{The Bad Apple Attack Applied to \bt{}}
\label{sec:bad-onion}

\begin{figure}[!t]
  \centering
  \includegraphics[width=0.6\columnwidth]{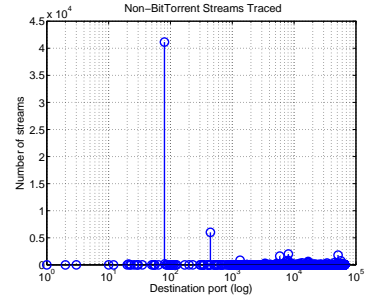}
  \caption{Number of non-\bt{} streams traced per destination port
    number.  \textit{The bad apple attack traces a significant number
      of HTTP (port 80) and HTTPS (port 443) streams.}}
  \label{fig:bad_onion}
 \end{figure}

Now that we have seen how to trace \bt{} streams, we describe the
specifics of the bad apple attack for \bt{} users.  As we have already
discussed, the bad apple attack can reveal the source IP address of
streams in the same circuit, or even different circuits.  With \bt{}
as with any other application, revealing the source IP address of
streams in the same circuit is straightforward; when the source IP
address of a \bt{} stream is revealed, all streams multiplexed into
the same circuit are associated with the same traced user.

To reveal the source IP address of streams in different circuits, we
exploit two patterns in \bt{} signalling traffic.  The first pattern
is the peer identifier which is essentially a random string of 20
bytes.  Thus, when we observe the same peer identifier in the \bt{}
messages of different circuits, we consider that these circuits have
the same source.  One problem with this first pattern is that it does
not work when communication between peers is encrypted.  To alleviate
this issue, we also consider communication to an IP/port freshly
returned in a tracker response as a second pattern.  In particular, if
in a circuit we observe that a peer initiates communication with an
IP/port contained in a tracker response from another circuit, we link
the two circuits.  We note that the linkage of \bt{} streams in
different circuits is particularly severe because, when an attacker
traces a Tor user, he can potentially associate past or future
circuits with that user, without need to reveal his IP address a
second time.

We found that the bad apple attack applied to \bt{} allowed us to
trace 193\% of additional streams as compared to BitTorrent streams,
including 27\% of HTTP streams possibly originating from ``secure''
browsers.  We show the number of non-BitTorrent streams traced per
destination port number in Fig.~\ref{fig:bad_onion}.

\begin{table*}
  \begin{minipage}{3in}
    \begin{center}
      \small
      \begin{tabular}{|c|c|c|c|c|}
        \hline
        Rank & \# & \% & Over & Country\\
        \hline
         1  & 958 & 14 & 0.9 & US\\
         2  & 937 & 13 & 5.6 & Japan\\
         3  & 887 & 13 & 2.8 & Germany\\
         4  & 369 & 5  & 1.3 & France\\
         5  & 354 & 5  & 1.8 & Poland\\
         6  & 236 & 3  & 0.9 & Italy\\
         7  & 232 & 3  & 0.6 & UK\\
         8  & 231 & 3  & - & China\\
         9  & 203 & 3  & 0.7 & Canada\\
         10 & 200 & 2  & 1.4 & Russia\\
        \hline
      \end{tabular}
  \end{center}
  \end{minipage}
  \begin{minipage}{2in}
    \begin{center}
      \small
      \begin{tabular}{|c|c|c|c|c|}
        \hline
        Rank & \# & Over & Country & AS\\
        \hline

        1  & 362 & 4.7   & Germany  & Deutsche Telekom (3320)\\
        2  & 274 & 5.7   & Japan    & NTT (4713)\\
        3  & 177 & 2     & Malaysia & TM Net (4788)\\
        4  & 142 & 1     & Italy    & Telecom Italia (3269)\\
        5  & 135 & 1.1   & France   & Orange (3215)\\
        6  & 133 & 1     & US       & AT\&T (7132)\\
        7  & 128 & 4.5   & Germany  & Hanse Net (13184)\\
        8  & 113 & - & China    & China Net (4134)\\
        9 & 109  & 1.4   & Poland   & TP Net (5617)\\
        10 & 104 & 1.8   & Austria  & UPC (6830)\\
        \hline
      \end{tabular}
    \end{center}
  \end{minipage}
  \caption{Popularity and over-representation of \bt{} users on Tor
    per country (left) and AS (right).}
  \label{tab:rank-top10-country}
\end{table*}

\section{Profiling Tor Users}
\label{sec:profil}

By analyzing the traffic relayed by our exit nodes, we evaluate that
19\% of all streams on Tor are BitTorrent streams.  (We remark that
this percentage is much higher than in McCoy et
al. \cite{Mccoy08shininglight}, suggesting that the number of
BitTorrent users on Tor has increased since 2008.)  We successfully
trace 9\% of \textit{all} Tor streams.  In this section, we use the
resulting 10,000 traced IP addresses to profile the BitTorrent
downloads and websites visited by Tor users per country of origin.

\subsection{BitTorrent Profiling}
\label{sec:bt_profiling}

We start by investigating whether BitTorrent utilization per country
and AS on Tor is different relatively to BitTorrent utilization
outside of Tor.  We then analyze the content downloaded by BitTorrent
users on Tor for a few interesting countries and investigate the
existence of an underground BitTorrent ecosystem on Tor.  Finally, we
analyze the websites visited by Tor users per country of origin.

\subsubsection{Utilization per Country of Origin}
\label{sec:country}

\begin{figure}[!t]
  \centering
  \includegraphics[width=0.6\columnwidth]{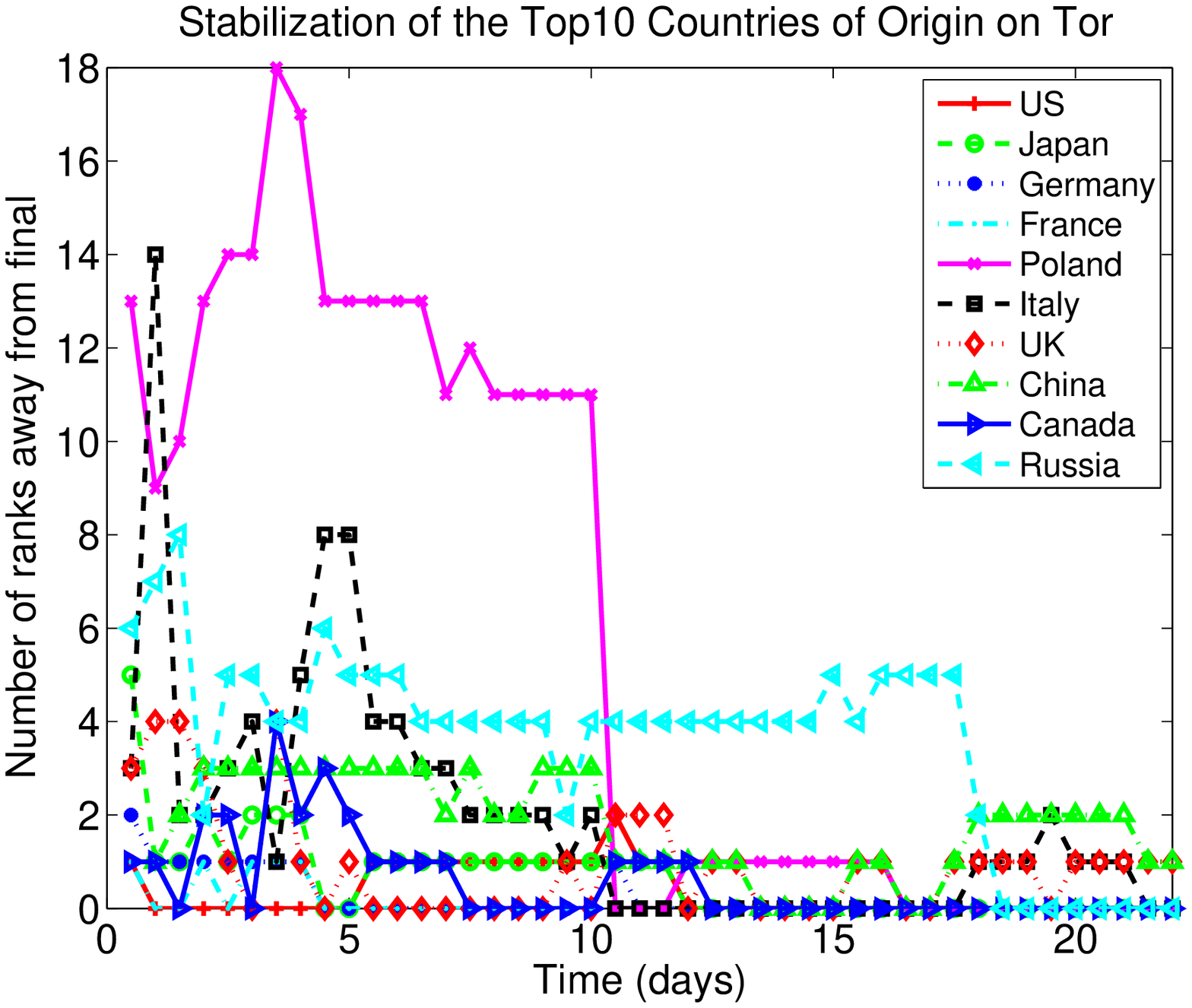}
  \includegraphics[width=0.6\columnwidth]{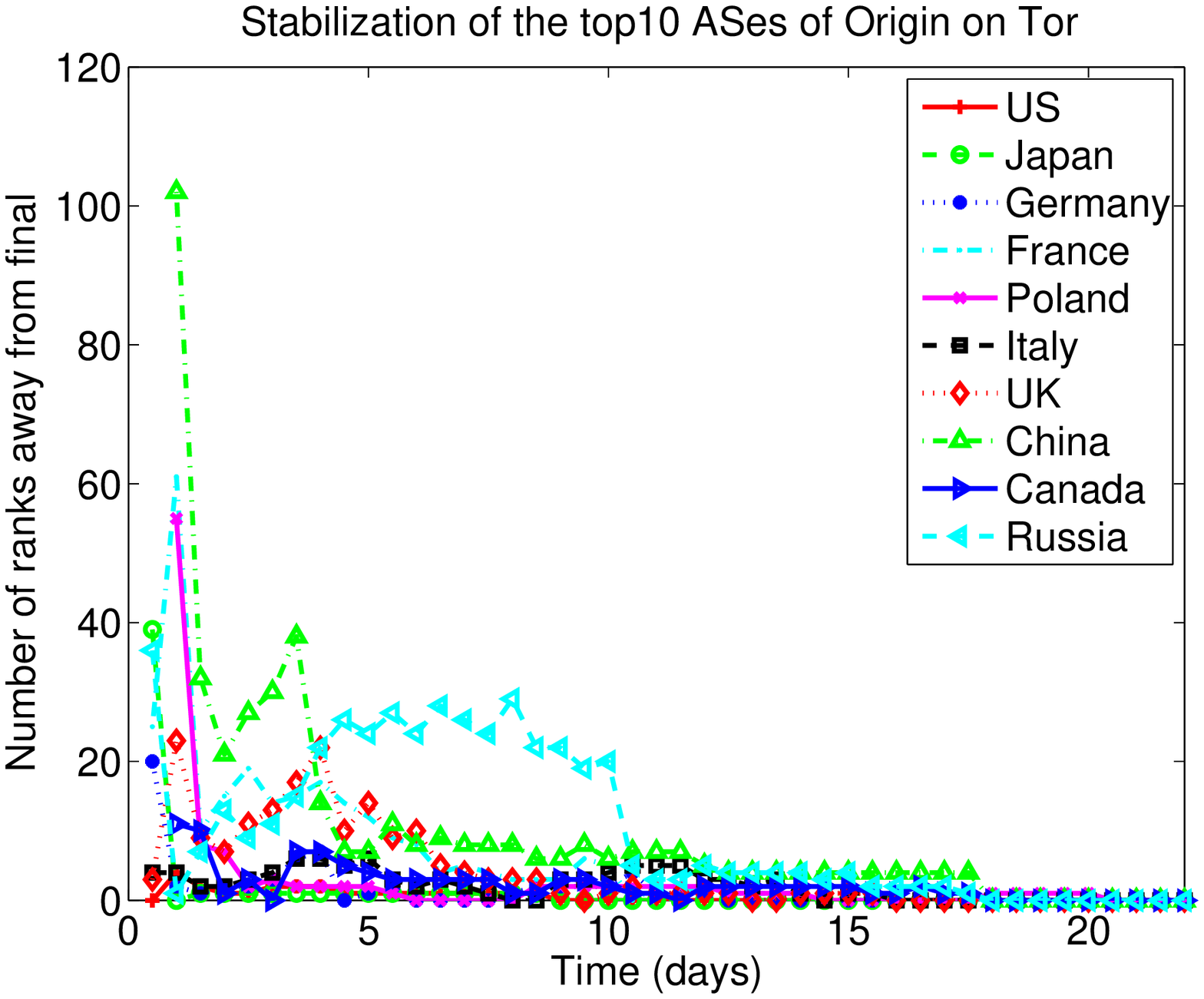}
  \caption{Stability of the top10 countries and ASes in cumulated
    number of IP addresses of traced Tor users.  For each country and
    AS, we plot the evolution of the difference between the current
    rank and the rank after 23 days. \textit{The duration of the
      measurement period is sufficient for the top10 countries and
      ASes to be reasonably representative of the overall utilization
      of BitTorrent on Tor.}}
  \label{fig:stability}
\end{figure}

To compare BitTorrent utilization on Tor and outside of Tor, we need a
representative sample of BitTorrent users for each of these two
utilizations.  We see in Fig.~\ref{fig:stability} that, after 10 days,
most of the top10 countries and AS (in number of IP addresses of
traced Tor users) have reached their final rank.  Therefore, we have a
reasonably representative sample of the utilization of BitTorrent {\it
  on Tor} in these countries and ASes.

As a representative sample of the utilization of BitTorrent {\it
  outside of Tor}, we then use a sample of 10,000,000 IP addresses
collected on August 22nd 2009 on the PirateBay, the largest BitTorrent
tracker at that time.  Indeed, the PirateBay was an order of magnitude
larger than the second largest BitTorrent tracker at the time of the
measurement \cite{BT-Public-Ecosystem} therefore we argue that a daily
sample from the PirateBay is reasonably representative of the global
utilization of BitTorrent outside of Tor.  We refer to Le Blond et
al. \cite{BT-Spying} for a description of the measurement methodology
to collect this data.

Table~\ref{tab:rank-top10-country} shows the popularity of \bt{} users
on Tor per country (left) and AS (right).  The over-representation
(Over) for a given country (resp. AS) is the fraction of \bt{} IP
addresses on Tor in that country (resp. AS) divided by the fraction of
IP addresses outside of Tor in the same country (resp. AS).  We do not
show the over-representation in China because Chinese content are not
generally tracked by the PirateBay, the tracker that we have used to
capture the location of BitTorrent users outside of Tor.  Hence, we
greatly over-estimate the over-representation in China.

An over-representation of $0.9$ in the US means that there is about
the same fraction of US \bt{} users on Tor as outside of Tor.  And an
over-representation of $5.6$ in Japan means that there are $5.6$ times
more \bt{} users from Japan on Tor than outside of Tor.  In other
words, whereas \bt{} US users do not hide on Tor more than average,
Japanese users strongly do.  The reasons behind such behavior may be
technological, political, sociological, etc.

\paragraph{Explaining Over-representations}
\label{sec:bt_downloads}

\begin{figure*}[!t]
  \includegraphics[height=0.82in]{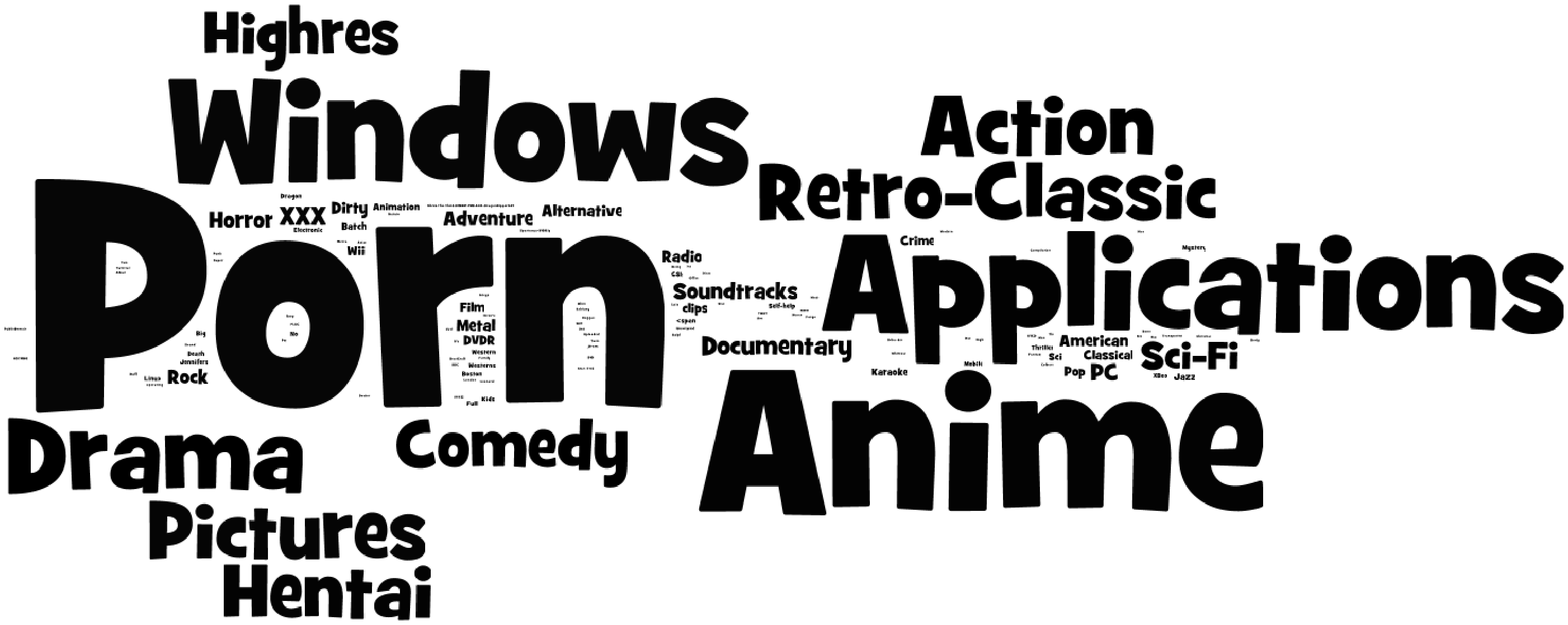}
  \includegraphics[height=0.82in]{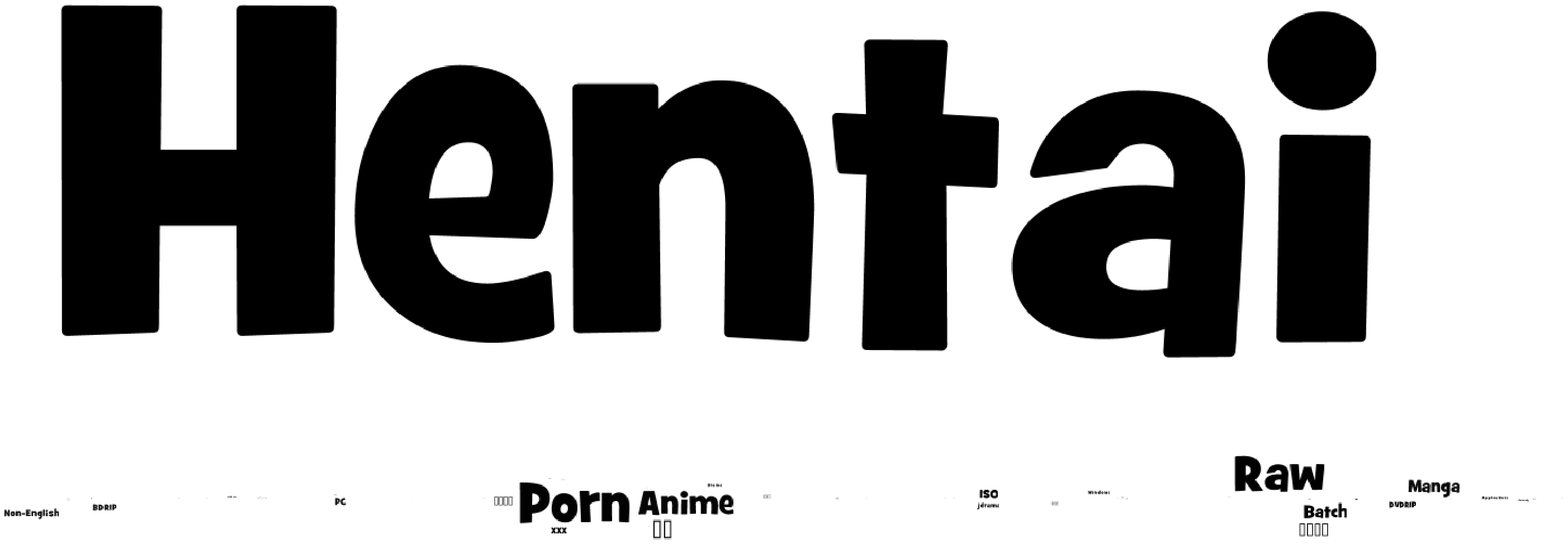}
  \includegraphics[height=0.82in]{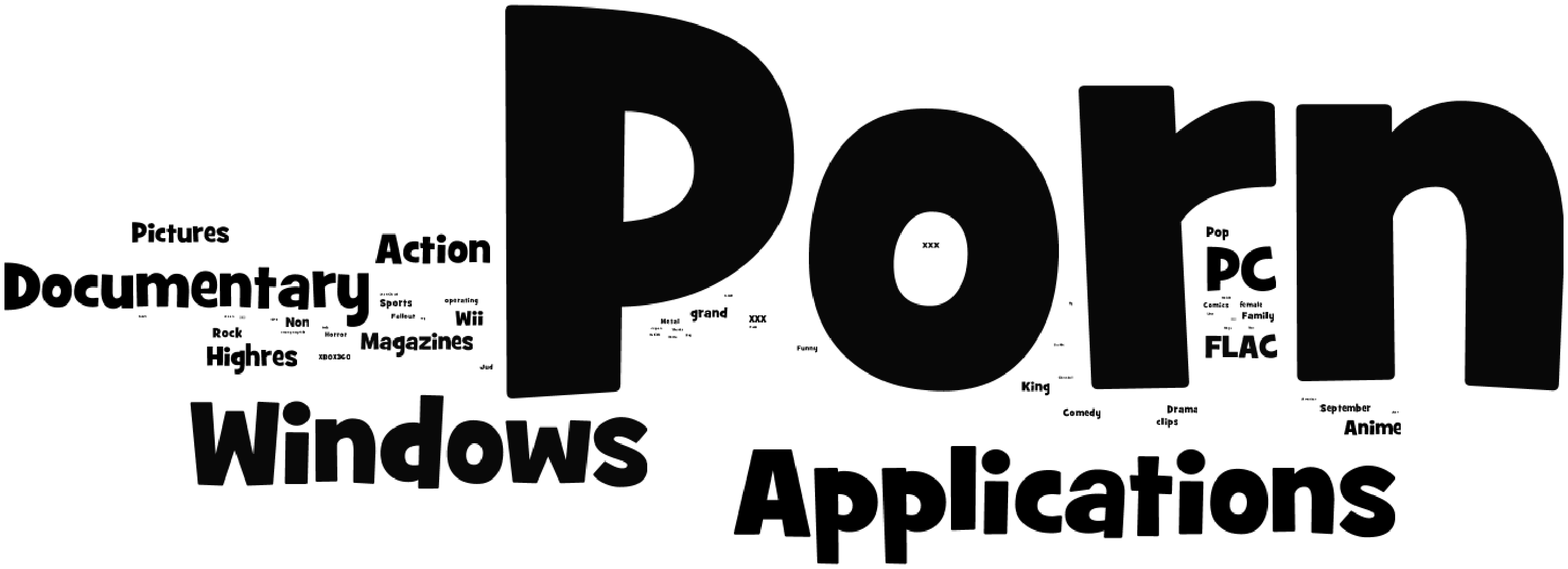}
  \caption{TagCloud of the US downloads (left), Japanese downloads
    (center), and German downloads (right).  We extract the tags of
    the BitTorrent content in the public ecosystem and vary the font
    size to reflect the number of content whose tag matches those
    keywords.  We increase the size of the keywords linearly with the
    frequency that they appear in the tags.  \textit{Japanese and
      German users use Tor to download much more pornographic material
      than US and other users.}}
  \label{fig:bt_profile}
\end{figure*}

To investigate the over-representations observed in
Section~\ref{sec:country}, we now analyze the types of content
downloaded by Tor users from countries with very different
over-representations.  In particular, we had observed that US users
were not over-represented on Tor whereas Japanese and German users
were.  In Fig.~\ref{fig:bt_profile}, we indeed see that US users are
downloading a large variety of content as compared to Japanese users
who mainly download Hentai (pornographic animes), and German users who
mainly download pornographic movies.  Therefore, we argue that the
reasons for over-representations (at least in BitTorrent) are mainly
sociological.

\subsubsection{The Underground Ecosystem}

\begin{figure}[!t]
  \centering
  \includegraphics[width=0.6\columnwidth]{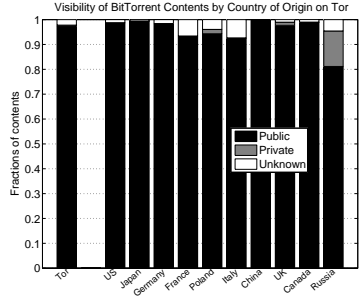}
  \caption{Distributions of BitTorrent content known in the
    \textit{Public} and \textit{Private} BitTorrent ecosystems, or
    \textit{Unknown} from the rest of the world.  \textit{A few
      content are unknown from the rest of the world.}}
  \label{fig:bt_public}
\end{figure}

BitTorrent comprises communities of users, or {\it ecosystems}, among
which content are distributed \cite{BT-Public-Ecosystem},
\cite{BT-Spying}, \cite{BT-Darknets}.  There is one {\it public
  ecosystem} that can be accessed by all Internet users
\cite{BT-Spying}, \cite{BT-Public-Ecosystem} and several {\it private
  ecosystems} with many users whose access is restricted to registered
users.  Even though private ecosystems are more difficult to monitor
than the public ecosystem, registration is generally open to everyone
and a single registered user can in principle monitor the whole
community \cite{BT-Darknets}.  Because even private ecosystems are
indeed relatively easy to monitor, it is probable that some private
ecosystems be known only by peculiar members (e.g., downloaders of
child pornography) thus forming an \textit{underground ecosystem}.

To investigate the existence of such an underground ecosystem on Tor,
we check whether there is a subset of the content distributed on Tor
that belongs neither to the public nor private ecosystems.  We use
.torrent files from 7,110,000 BitTorrent content: 6,800,000 from the
public BitTorrent ecosystem \cite{BT-Spying, BT-Public-Ecosystem}, and
310,000 from the private BitTorrent ecosystem \cite{BT-Darknets}.  We
also search the missing .torrent files on Google.  To the best of our
knowledge, this is the largest collection of .torrent files ever
assembled.  We see in Fig.~\ref{fig:bt_public} that 3\% of all the
content distributed by BitTorrent users on Tor belong neither to the
public nor private ecosystems.  This result suggests the existence of
an underground ecosystem on Tor.  However, one would need to download
these unknown content and to check them manually in order to determine
whether they belong to regular private ecosystems for which we do not
have the .torrent files or whether they are more sensitive.

\subsection{Web Profiling}

\begin{figure}[!t]
  \centering
  \includegraphics[ width=0.6\columnwidth]{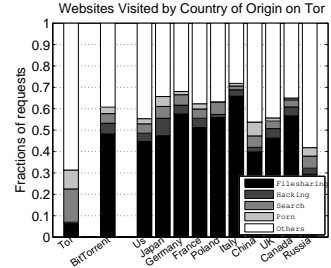}
  \caption{Distributions of the categories of websites visited by Tor
    users per country of origin. {\it Tor} represents the overall
    distribution of requests for all Tor users (not necessarily
    traced). {\it BitTorrent} represents the overall distribution of
    requests for traced Tor users. The other bars represent the same
    information but per country of origin. \textit{BitTorrent users on
      Tor visit significantly more Hacking websites and significantly
      less Search and Porn than regular Tor users.}}
  \label{fig:http_usage}
\end{figure}

We have showed that the bad apple attack allowed us to trace 193\% of
additional streams as compared to BitTorrent streams, including 27\%
of HTTP streams possibly originating from ``secure'' (i.e., proxied)
browsers.  In the following, we use the web-filtering service of
FortiGuard \cite{fortiguard} to analyze the type of websites visited
on Tor per country of origin.

We show the type of websites visited per country of origin in
Fig.~\ref{fig:http_usage}.  Because they all are BitTorrent users,
traced users differ from the average Tor users (\textit{Tor}).  In
particular, around 50\% of the requests are targeted to filesharing
websites (\textit{FileSharing}) such as \textit{ThePirateBay},
\textit{MegaDownload}, or \textit{RapidShare}.  Traced Tor users also
visit significantly more hacking websites suggesting that they are
interested in security.  Finally, they also visit significantly less
search and porn websites than average Tor users.  This might be
because they already rely on BitTorrent rather than the Web to search
and download content, including pornographic material.

\section{Related Work}
\label{sec:related}

Tor's efficiency has required several adaptations to the original
design of onion routing \cite{Reed98anonymousconnections} whose impact
on privacy are not well understood in practice \cite{Tor}.  In
particular, we showed that the multiplexing of several streams into a
single circuit can significantly degrade privacy.

To date, Tor measurement studies and attacks have been carried out in
isolation.  Measurement studies have documented {\it who is using Tor}
and {\it how Tor is used} but without the ability of associating the
two information, e.g., to profile Tor users
\cite{Mccoy08shininglight}.  Attacks have documented methodologies to
associate the two information but without actually profiling Tor users
\cite{TOR-Hacking, TOR-Securing, TOR-Browser}.  This paper strikes a
balance between the two by developing new attacks targeted to P2P
applications, launching these attacks at a reasonable scale against
the Tor network, and profiling Tor users.

\subsection{Web-level Tracing Attacks}

For simplicity, we assume that the attacker always controls an exit
node in the attacks described hereafter.

We now describe attacks targeting Web applications to reveal the IP
address of Tor users.  FortConsult designed two attacks based on
active content injection by the exit nodes to trace Tor users
\cite{TOR-Hacking}.  The first attacks consisted in Flash injection so
that the targeted user connects to a server controlled by the attacker
{\it outside} of Tor, hence exposing his IP address.  (A cookie is
used to associate that IP address to the stream in Tor.)  The second
attack consisted in injecting JavaScript to send the (local) IP
address of the user over Tor.  This study reported that whereas the
first attack was effective, the second was not mainly because local IP
addresses sent over Tor were not routable, e.g., 192.168.0.1.

Abbott et al. relied on JavaScript and HTML meta refresh tag to inject
timing patterns \cite{TOR-Browser}.  The assumption being that users
will leave that page open long enough so that pattern can be spotted
by an entry node, also controlled by the attacker (thus tracing the
user).  It is interesting to note that even users having disabled
active content are susceptible to HTML meta refresh tag injection.

To the best of our knowledge, we are the first to design attacks
against P2P applications on Tor, to validate these attacks at a
reasonable scale, and to demonstrate that one can associate many
streams possibly originating from ``secure'' (i.e., proxied) browsers,
with traced users.  We remark that even though we have targeted P2P
applications in this paper, the bad apple attack can originate from
any insecure application.

\subsection{Measurements Studies}

The main measurement study of Tor that we are aware of has been made
by McCoy et al. \cite{Mccoy08shininglight}.  The authors provided
interesting insights into {\it who is using Tor} and {\it how Tor is
used and mis-used}.  In particular, they showed that BitTorrent
generates 40\% of all traffic on Tor and claimed that it represents
only 3\% of all streams.

We complemented this measurement study in two important aspects.
First, we showed that 70\% of BitTorrent users establish P2P
connections {\it outside} of Tor thus making most BitTorrent streams
(and traffic) invisible to the Tor network.  We argue that this
finding makes of BitTorrent users a target of choice on Tor.  Second,
we launched attacks to profile Tor users.  This profiling brought
elements of answer to one important question raised by McCoy et al.,
that is, ``... why there is such a large scale adoption of Tor in
[...]  specific countries, relative to Tor usage in other countries.''
We showed that the answer to that question (at least for BitTorrent)
is unlikely to be technological or political but in fact sociological.

\section{Summary and Perspectives}
\label{sec:discuss}

Using BitTorrent as an insecure application, we designed two attacks,
one consisting in hijacking tracker responses and one exploiting the
statistical properties of the DHT, to trace BitTorrent streams on Tor.
We then showed that the bad apple attack allows us to trace
non-BitTorrent streams.  In particular, we traced 193\% of additional
streams as compared to BitTorrent streams, including 27\% of HTTP
streams possibly originating from ``secure'' (i.e., proxied) browsers.
In total, we traced 9\% of \textit{all} Tor streams carried by our
instrumented exit nodes.

\vskip 0.2cm

We ran these attacks in the wild for 23 days and reveal 10,000 IP
addresses of Tor users.  Using these IP addresses, we then profiled
not only the BitTorrent downloads but also the websites visited per
country of origin of Tor users.  In particular, we showed that
sociological reasons could explain the large number of Tor users in
certain countries relatively to other.  Finally, we presented results
that suggest the existence of an underground \bt{} ecosystem on Tor.

\vskip 0.8cm

Defending against the bad apple attack is not straightforward.  The
most effective defense would be to have one stream per circuit, as in
the original onion routing, however, performances issues make this
defense unfeasible.  Another defense would be to isolate streams by
groups of destination port in different circuits, e.g., the secure and
the insecure circuit.  Destination ports known to be used by secure
applications, e.g., 80, 22, would use the secure circuit thus limiting
the risk that the source IP address of one stream in that circuit gets
revealed by an insecure application.  One weakness of this defense is
that an attacker could trick an insecure application into connecting
to a port that is usually used by a secure application, thus
multiplexing the insecure stream into the secure circuit.  Yet another
defense would be to isolate each application into its own circuit,
hence compartmenting the bad apple attack to the insecure application.
However, modern operating systems lack a portable way to map an
incoming stream to an application.  We have discussed our results and
possible solutions to address the bad apple attack with the Tor
project.

\vskip 0.8cm

We remark that even though the bad apple attack does not exist in
application-level anonymity networks dedicated to a single application
(e.g., OneSwarm \cite{Oneswarm}), the corpus of networking
applications is too broad to practically build one network for each
application.  In this respect, we believe that we have validated an
important attack against the design of modern anonymity networks and
that we should defend against it to protect users privacy on the
Internet.

\vskip 0.8cm

{\bf Acknowledgements} We warmly thank Keith Ross and Chao Zhang for
sharing their identifiers of public and private content with us.  We
are also grateful to Roger Dingledine, Nick Mathewson, the anonymous
reviewers, and our shepherd, Thorsten Holz, for their constructive
comments.

\begin{small}

\end{small}


\begin{thebibliography}{10}

\bibitem{TOR-Browser}
T.~Abbott, K.~Lai, M.~Lieberman, and E.~Price.
\newblock {Browser-based Attacks on Tor}.
\newblock In {\em Proc. of PETS'07}, Ottawa, Canada, 2007.

\bibitem{BT-Spying}
S.~L. Blond, A.~Legout, F.~Lefessant, W.~Dabbous, and M.~A. Kaafar.
\newblock {Spying the World from your Laptop - Identifying and Profiling
  Content Providers and Big Downloaders in BitTorrent}.
\newblock In {\em Proc. of LEET}, San Jose, CA, USA, 2010.

\bibitem{TOR-BT-Poster}
S.~L. Blond, P.~Manil, A.~Chaabane, M.~A. Kaafar, C.~Castelluccia, A.~Legout,
  and W.~Dabbous.
\newblock {De-anonymizing BitTorrent Users on Tor}.
\newblock In {\em Proc. of NSDI'10, Poster session}, San Jose, CA, 2010.

\bibitem{TOR-Blog}
R.~Dingledine.
\newblock {Bittorrent Over Tor isn't a Good Idea}.
\newblock
  \url{https://blog.torproject.org/blog/bittorrent-over-tor-isnt-good-idea},
  2010.

\bibitem{Tor}
R.~Dingledine, N.~Mathewson, and P.~Syverson.
\newblock {Tor: the Second-generation Onion Router}.
\newblock In {\em Proc. of USENIX}, Boston, MA, 2004.

\bibitem{TOR-Hacking}
FortConsult.
\newblock {Practical Onion Hacking: Find the Real Address of Tor Clients}.
\newblock
  \url{http:www.fortconsult.net/images/pdf/Practical_Onion_Hacking.pdf}, 2006.

\bibitem{fortiguard}
FortiGuard.
\newblock Web filtering.
\newblock \url{http://www.fortiguard.com/webfiltering/webfiltering.html}.

\bibitem{Oneswarm}
T.~Isdal, M.~Piatek, A.~Krishnamurthy, and T.~Anderson.
\newblock {Privacy-Preserving P2P Data Sharing With OneSwarm}.
\newblock In {\em Proc. of SIGCOMM}, Bangalore, India, 2010.

\bibitem{loesing:financialcrypto2010}
K.~Loesing, S.~Murdoch, and R.~Dingledine.
\newblock {A Case Study on Measuring Statistical Data in the Tor Anonymity
  Network}.
\newblock In {\em Proc. of Financial Cryptography and Data Security '10}, 2010.

\bibitem{TOR-BT-TR}
P.~Manils, A.~Chaabane, S.~L. Blond, M.~A. Kaafar, C.~Castelluccia, A.~Legout,
  and W.~Dabbous.
\newblock {Compromising Tor Anonymity Exploiting P2P Information Leakage}.
\newblock Technical report, {INRIA}, {2010}.

\bibitem{Mccoy08shininglight}
D.~Mccoy, T.~Kohno, and D.~Sicker.
\newblock {Shining Light in Dark Places: Understanding the Tor Network}.
\newblock In {\em Proc. of PETS'08}, Leuven, Belgium, 2008.

\bibitem{TOR-Securing}
M.~Perry.
\newblock {Securing the Tor Network}.
\newblock In {\em Proc. of Black Hat}, Las Vegas, NV, 2007.

\bibitem{BT-DMCA}
M.~Piatek, T.~Kohno, and A.~Krishnamurthy.
\newblock {Challenges and Directions for Monitoring P2P File Sharing Networks
  or Why My Printer Received a DMCA Takedown Notice}.
\newblock In {\em Proc. of HotSec}, San Jose, CA, 2008.

\bibitem{Reed98anonymousconnections}
M.~G. Reed, P.~F. Syverson, and D.~M. Goldschlag.
\newblock {Anonymous Connections and Onion Routing}.
\newblock {\em IEEE Journal on Selected Areas in Communications}, 16:482--494,
  1998.

\bibitem{BT-Darknets}
C.~Zhang, P.~Dhungel, Z.~Liu, and K.~W. Ross.
\newblock {BitTorrent Darknets}.
\newblock In {\em Proc. of INFOCOM}, San Jose, CA, USA, 2010.

\bibitem{BT-Public-Ecosystem}
C.~Zhang, P.~Dhungel, D.~Wu, and K.~W. Ross.
\newblock {Unraveling the BitTorrent Ecosystem}.
\newblock {\em TPDS}, 2010.

\end{thebibliography}
\end{document}